\documentclass[preprint,aps,showpacs,showkeys,dvips]{revtex4}

\usepackage{graphicx}
\usepackage{dcolumn}
\usepackage{bm}
\usepackage{subfigure}

\begin{document}

\title{A Comperative Study of the Anomalous Single Production of the Fourth Generation Quarks at $ep$ and $\gamma p$ Colliders}

\author{R. \c{C}ift\c{c}i}
\email{rena.ciftci@cern.ch}
\affiliation{A. Yuksel Cad. 7/15, 06010 Etlik, Ankara, Turkey}
\author{A. K. \c{C}ift\c{c}i}
\email{ciftci@science.ankara.edu.tr}
\affiliation{Physics Department, Faculty of Sciences, Ankara University, 06100 Tandogan,
Ankara, Turkey} 


\begin{abstract}
We propose some channels for the possible observation of anomalous interactions of the fourth standard model generation quarks at the Large Hadron Collider based $ep$ and $\gamma p$ colliders. Namely, $u_4 (d_4) \rightarrow q \gamma $ and $u_4 (d_4) \rightarrow q Z \rightarrow q \ell^+ \ell^- $ decay processes are considered. Signatures for signals and corresponding standard model backgrounds are investigated at both colliders comperatively. The lowest necessary luminosities to observe these processes and the achievable values of the anomalous coupling strengths are determined. It is shown that the $\gamma p$ collider is advantageous compare to the $ep$ collider.
\end{abstract}
\maketitle
\section{Introduction}	

As well known, the number of the fermion generations is not determined by the Standard Model (SM). LEPI data obtained at the Z pole have established that the number of generations with light neutrinos is equal to three \cite{Yao06}. However, the data do not limit the number of generations with heavy neutrinos ($m_\nu =m_Z /2$). Then again, the upper limit for this number is determined to be less than nine by the asymptotic freedom in QCD.

Also, SM does not explain the mass and mixing values of the fundamental fermions. Based on the naturalness arguments the flavor democracy is proposed as a procedure to give an explanation for these (see review  \cite{Ciftci05} and references therein). The existence of the fourth SM generation is an unavoidable outcome of  this approach \cite{Fritzsch87,Datta93,Celikel95}. In addition, the recent precision electroweak data are equally consistent with the presence of three or four fermion SM generations \cite{He01,Novikov02,Kribs07}. Meanwhile, there are phenomenological arguments against existence of fifth SM generation  \cite{Sultansoy00}. The masses of the fourth generation fermions are predicted by the flavor democracy to be quasi degenerate and lie between 300 and 700 GeV. Experimentally, the lower bound on the mass of the fourth generation quarks is given by CDF at Tevatron as $256$ GeV \cite{CDF}. 

The fourth generation quarks will be produced in pairs copiously at the Large Hadron Collider (LHC) \cite{Arik99,ATLAS99,Holdom06,Holdom07a,Holdom07b,Ozcan08}. Even though the discovery of the fourth generation neutrinos at the LHC might be somewhat possible through pair production \cite{Cuhadar08},  it will not be good place to observe the fourth generation charged leptons through pair production.   Lepton colliders are the best place for pair production of the fourth generation charged lepton and neutrino \cite{Ciftci05}. Also, lepton colliders are good place to look at the fourth family quarkonia formation \cite{Ciftci03}. Moreover, Ref. \cite{Fritzsch99} reasons that t-quark anomalous interactions may exist due to its large mass. With similar arguments, one can show that anomalous interactions appear to be quite natural for the fourth generation fermions.  The discovery capacity of lepton collider could be enlarged if the anomalous interactions of the fourth generation fermions with the first three ones exist.  These anomalous interactions could provide also single production of the fourth generation fermions at future lepton-hadron \cite{Ciftci08a,Ciftci08b} and gamma-hadron colliders. Lepton-hadron colliders with $\sqrt{s}$ = 1.3-1.8 TeV  are named QCD Explorer or LHeC depending on electrons provided by linac or ring, respectively \cite{Sultansoy04}. Also, it is possible to build a $\gamma p$ collider on the base of $ep$ collider by using Compton backscattering technique  \cite{Ciftci95,Ciftci98,Aksakal07,Ciftci07}.   In this paper, a comperative study of the possible anomalous single productions of the fourth generation quarks are made at LHC based $ep$ and $\gamma p$ colliders.

\begin{figure}[b]
\vspace{-0.2cm}
\subfigure[]{\includegraphics[width=7.0cm]{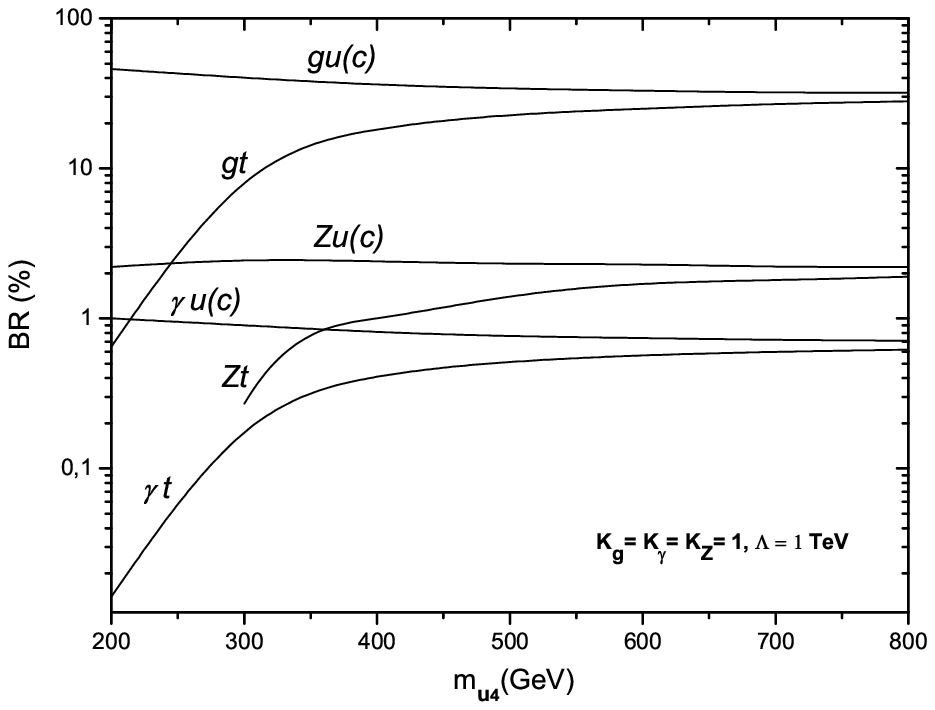}}\ \ \ \ \ \ \ \ \ \ 
\subfigure[]{\includegraphics[width=7.0cm]{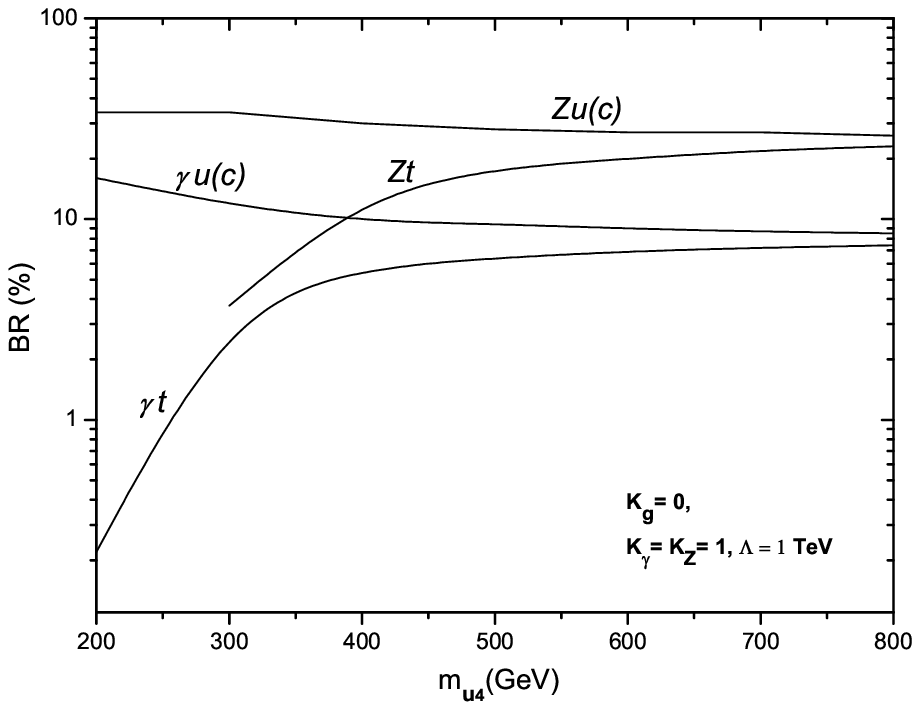}}\ \ \ \ \ \ \ \ \ \ 
\subfigure[]{\includegraphics[width=7.0cm]{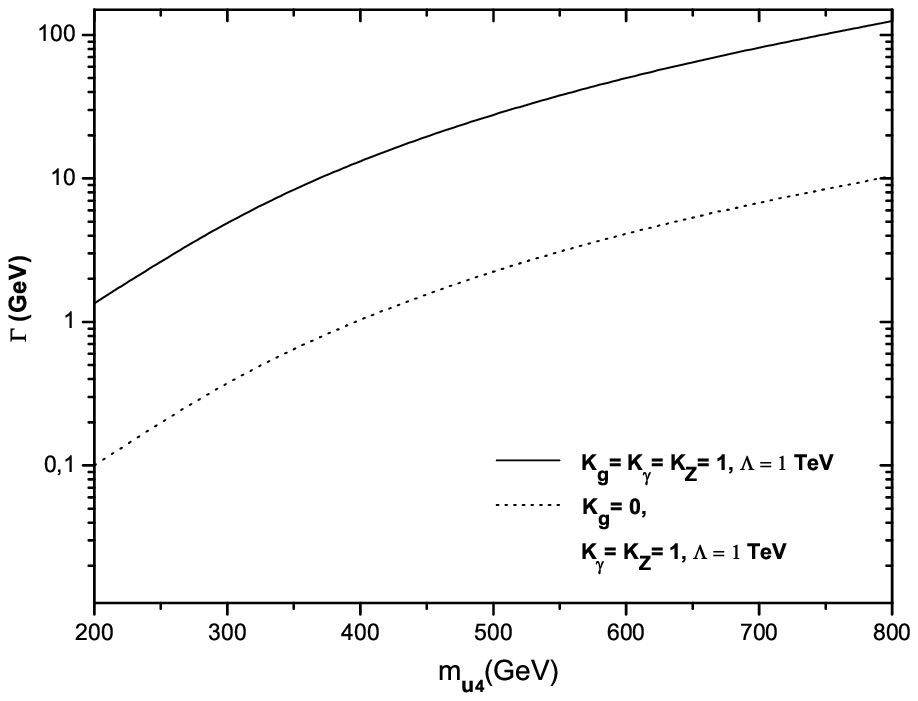}}
\vspace{-0.2cm}
\caption{(a), (b) branching ratios and (c) the total decay width of the fourth SM generation up type quark as a function of the quark mass. \protect\label{fig1}}
\end{figure}

Detailed argumentations of the LHC based $ep$ and $\gamma p$ colliders are given in Section II. In Section III, the Lagrangian for anomalous interactions of the fourth generation quarks is presented; the decay width and branching ratios of the fourth generation quarks are evaluated. The possible single anomalous productions of the fourth generation quarks at LHC based $ep$ and $\gamma p$ colliders are studied in Section IV: $u_4 (d_4) \rightarrow q \gamma $ and $u_4 (d_4) \rightarrow q Z \rightarrow q \ell^+ \ell^- $ decay processes at both colliders are considered as a signature of anomalous interactions of the fourth generation quarks (where $q$ is $u$ and $c$ for $u_4$ and $d,s$ and $b$ for $d_4$ and $\ell$ is $e$ and $\mu$) as well as their SM backgrounds. And, statistical significances of signals are compared at both colliders with respect to production of processes given above. The lowest necessary luminosities to observe these processes and the achievable values of the anomalous coupling strengths are given in Section V.

\begin{figure}[b]
\vspace{-0.2cm}
\subfigure[]{\includegraphics[width=7.0cm]{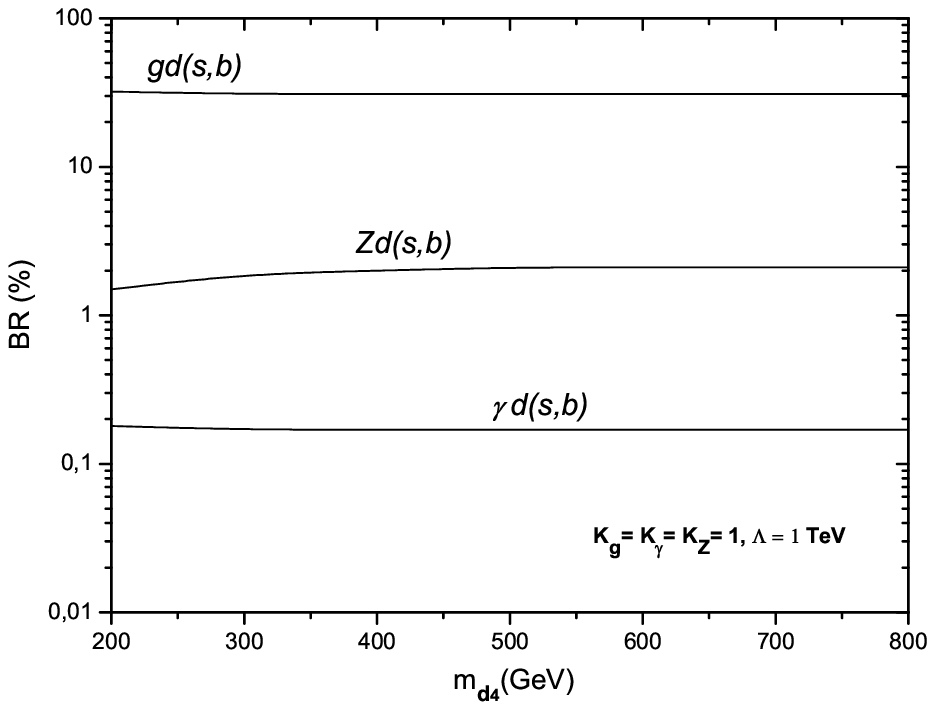}}\ \ \ \ \ \ \ \ \ \ 
\subfigure[]{\includegraphics[width=7.0cm]{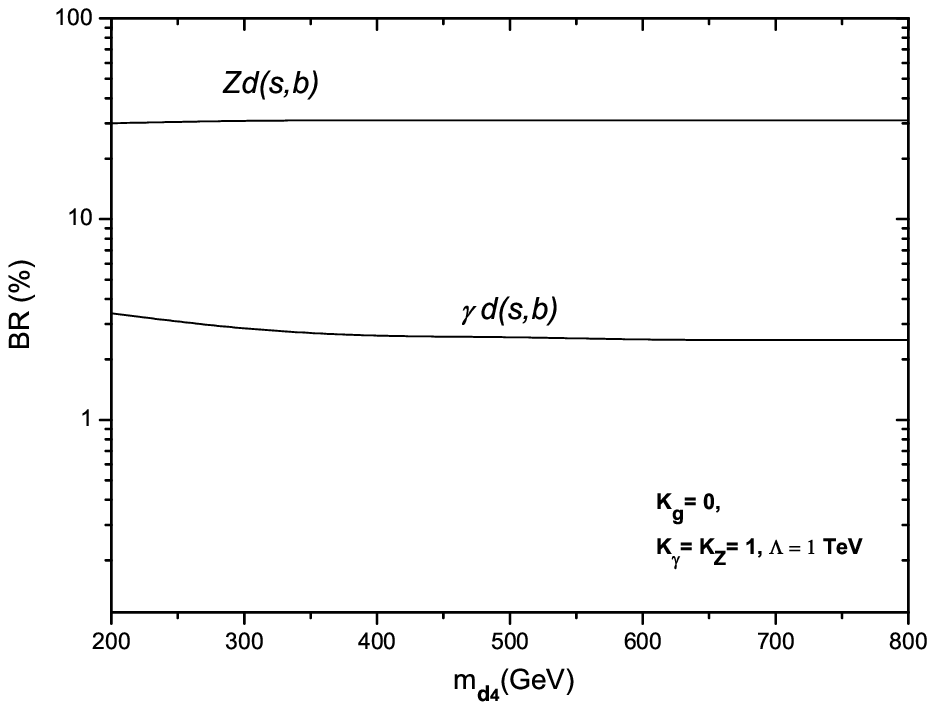}}\ \ \ \ \ \ \ \ \ \ 
\subfigure[]{\includegraphics[width=7.0cm]{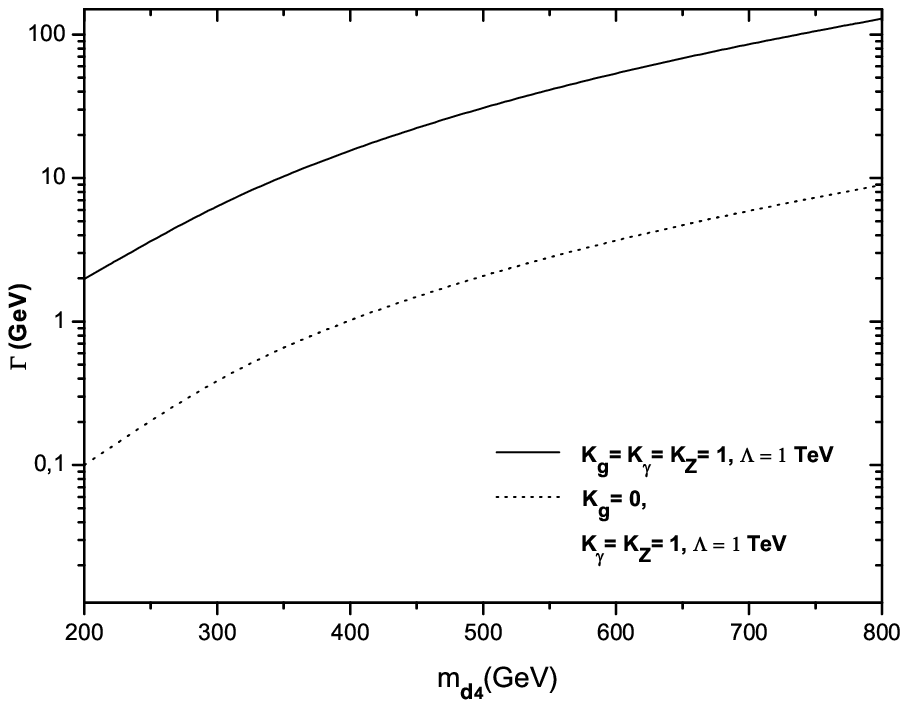}}
\vspace{-0.2cm}
\caption{(a), (b) branching ratios and (c) the total decay width of the fourth SM generation down type quark as a function of the quark mass. \protect\label{fig2}}
\end{figure}

\section{LHC based $ep$ and $\gamma p$ Colliders}

The LHC based $ep$ collider is proposed to extend the discovery reach of the hadron collider and enable the precision physics with the LHC  \cite{Sultansoy04}. To reach required center of mass energies to study QCD, the LHC 7 TeV proton beam should collide with 60-120 GeV electron beam. For our purposes in this paper, the electron beam energy is chosen as 60 GeV. Either a linac or e-ring at LHC tunnel can be used for electron beam. 

For minimal interuption of LHC run, the linac alternative is selected in this paper. Estimations of the luminosity values for various configuration and parameter sets of the LHC and super conducting (s.c.) electron accelerator are given in \cite{Zimmermann}. Integrated luminosities for these accelerator parameter sets vary from  $0.05 fb^{-1}$ to $10 fb^{-1}$. These luminosities can be achieved by using either a pulsed linac at a total electric wall plug power of 100 MW or continuous wave (cw) mode linac with the option of energy recovery at a total electric wall plug power of 20 MW. 

For the choice of a cw mode superconducting energy recovery e-linac with an upgraded LHC, the integrated luminosity for a year has been estimated $10  fb^{-1}$. However, for the choice of pulsed superconducting e-linac with an upgraded LHC, the integrated luminosity per year has been determined as $4.1 fb^{-1}$. While the first choice has a better luminosity, it is not possible to found a gamma-p collider on base of it. The real gamma production through the Compton backscattering technique on the base of the ep collider is possible on the second choice. In this case, the integrated luminosity of $\gamma p$ collider reaches the maximum value of $2.665 fb^{-1}$.

\begin{figure}[t]
\vspace{-0.2cm}
\subfigure[]{\includegraphics[width=7.9cm]{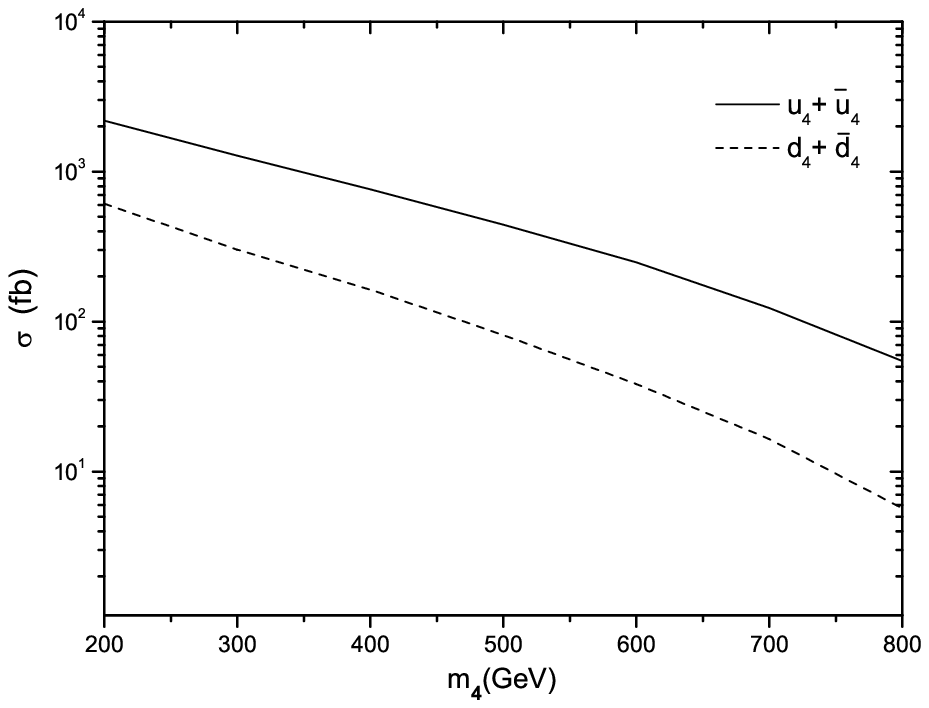}}\ \    
\subfigure[]{\includegraphics[width=7.9cm]{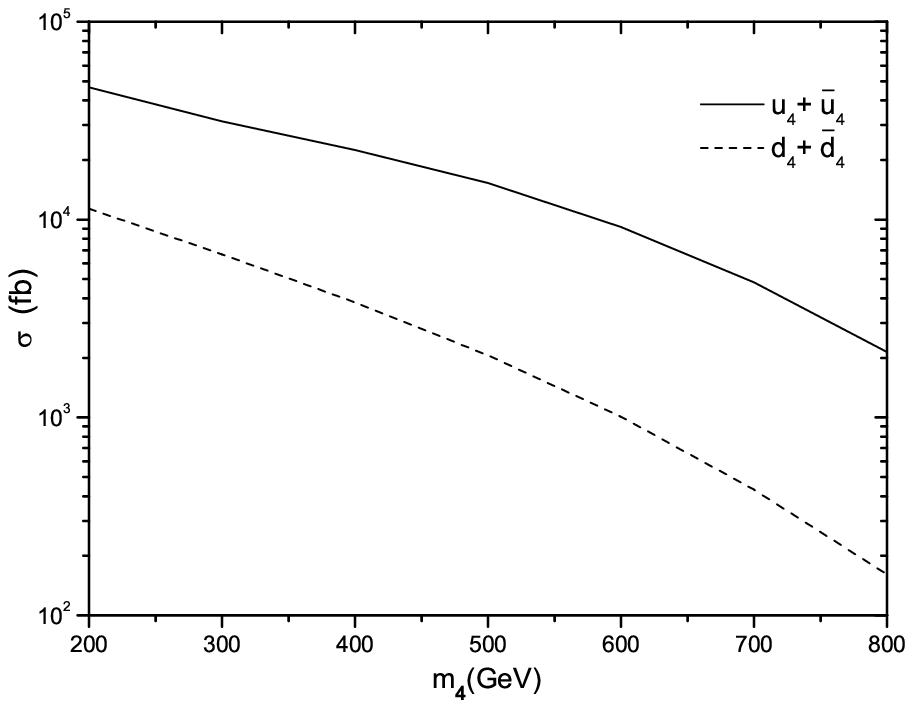}}
\vspace{-0.2cm}
\caption{The total production cross sections of the fourth SM generation up and down type quarks at (a) $ep$ colliders and (b) $\gamma p$ colliders. \protect\label{fig3}}
\end{figure} 

\section{Anomalous Single Production of the Fourth Generation Quarks at future $ep$ and  $\gamma p$ colliders}

The effective Lagrangian for the flavor changing neutral currents  (FCNC) interactions of $u_4$ and $d_4$ quarks can be rewritten from \cite{Rizzo97,Arik03b} with minor modifications as:
\begin{equation}
{\cal L}=\left(\frac{\kappa^{q_{i}}_{\gamma}}{\Lambda}\right) e_{q}g_{e}\bar{q}_{4}\sigma_{\mu\nu}q_{i} F^{\mu\nu}+\left(\frac{\kappa^{q_{i}}_{Z}}{2\Lambda}\right) g_{Z}\bar{q}_{4}\sigma_{\mu\nu}q_{i} Z^{\mu\nu}+\left(\frac{\kappa^{q_{i}}_{g}}{\Lambda}\right) g_{s}\bar{q}_{4}\sigma_{\mu\nu}T^{a}q_{i} G^{\mu\nu}_{a}+h.c. \hspace{2mm}  ,  \hspace{3mm} 
\end{equation}
where $\textit{i}=1,2,3$ denotes the generation index. $\kappa^{q_{i}}_{\gamma}$, $\kappa^{q_{i}}_{Z}$ and $\kappa^{q_{i}}_{g}$  are anomalous couplings for the electromagnetic, the weak (neutral current) and the strong interactions, respectively (in numerical calculations, $\kappa^{q_{i}}_{\gamma}= \kappa^{q_{i}}_{Z} = \kappa^{q_{i}}_{g}$ is assumed). $\Lambda$ is the cutoff scale for the new physics and $e_q$ is the quark charge. $g_e$, $g_Z$ and $g_s$ are the electroweak and the strong coupling constants. In the above equation, $\sigma_{\mu\nu} = i(\gamma_{\mu}\gamma_{\nu}-\gamma_{\nu}\gamma_{\mu})/2$. $F^{\mu\nu}$, $Z^{\mu\nu}$ and $G^{\mu\nu}_{a}$ are field strength tensors of the photon, the Z boson and gluons, respectively. $T_a$ is the Gell-Mann matrices.

We have calculated the anomalous single production cross sections of the fourth SM generation quarks at the linac-LHC $e p$ colliders and $\gamma p$ colliders based on it using CompHEP with CTEQ6L1 \cite{Pumplin02}. The total decay width $\Gamma$ of the fourth generation up (down) type quarks  and the relative branching ratios are plotted with assumption of $\kappa^{q_{i}}_{\gamma}= \kappa^{q_{i}}_{Z} = \kappa^{q_{i}}_{g}=1$  ($\kappa^{q_{i}}_{\gamma}= \kappa^{q_{i}}_{Z} =1, \; \kappa^{q_{i}}_{g}=0$) in Fig. 1 (Fig. 2) with $\Lambda=1$TeV. At the rest of the study, the anomalous coupling for the strong interactions is chosen equal to zero. Single anomalous production cross sections of the fourth generation quarks are given for ep collider at Fig. 3a and $\gamma p$ collider at Fig. 3b. 

\begin{table}[b]
\caption{Signal and SM background cross sections for $q_4 \rightarrow \gamma q$ anomalous process at $ep$ and $\gamma p$ colliders with $\kappa_{g}=0$, $(\kappa_{\gamma}/\Lambda) = (\kappa_{Z}/\Lambda)=1$ TeV$^{-1}$.}   
\begin{ruledtabular}
\begin{tabular}{ccccc}
 &\multicolumn{4}{c} {Signal cross sections (fb)} \\
 & \multicolumn{2}{c}{$e p \rightarrow q_4 X\rightarrow e \gamma j X+h.c.$}&\multicolumn{2}{c}{$\gamma p \rightarrow q_4 X\rightarrow \gamma j  X+h.c.$}   \\ 
 \cline{2-3} \cline{4-5}
 $m_{q_{4}}$ (GeV)&Selection 1&Selection 2&Selection 1&Selection 2\\
\hline
200 & 196.60 & 138.50 & 10850 & 7610 \\
300 & 90.60 & 86.86 & 5410 & 5330 \\
400 & 45.71 & 45.51 & 3035 & 3035  \\
500 & 22.76 & 22.76 & 1672 & 1672 \\
600 & 10.86 & 10.86 & 838 & 838 \\
700 & 4.69 & 4.69 & 374 & 374 \\
800 & 1.72 & 1.72 & 124 & 124 \\
\hline
SM Backg. (fb)&13080 & 120 & 90200 & 1220 \\
\end{tabular}
\end{ruledtabular}
\end{table}

\begin{table}[b]
\caption{Signal and SM background cross sections for $q_4 \rightarrow Z q$ anomalous process at $ep$ and $\gamma p$ colliders with $\kappa_{g}=0$, $(\kappa_{\gamma}/\Lambda) = (\kappa_{Z}/\Lambda)=1$ TeV$^{-1}$.}  
\begin{ruledtabular}
\begin{tabular}{ccccc}
 &\multicolumn{4}{c} {Signal cross sections (fb)} \\
 & \multicolumn{2}{c}{$e p \rightarrow q_4 X\rightarrow e \ell^+ \ell^- j X+h.c.$ }&\multicolumn{2}{c}{$\gamma p \rightarrow q_4 X\rightarrow  \ell^+ \ell^- j X+h.c.$}  \\ 
 \cline{2-3} \cline{4-5}
 $m_{q_{4}}$ (GeV)&Selection 1&Selection 2&Selection 1&Selection 2\\
\hline
200 & 32.20 & 4.61 & 1786 & 92 \\
300 & 15.96 & 5.12 & 1019 & 266 \\
400 & 8.87 & 3.96 & 568 & 249 \\
500 & 4.20 & 2.42 & 302 & 168 \\
600 & 1.62 & 1.29 & 148 & 95 \\
700 & 0.86 & 0.59 & 61 & 43 \\
800 & 0.30 & 0.22 & 19 & 14\\
\hline
SM Backg. (fb)& 19.24 & 0.64 & 2960 & 44 \\
\end{tabular}
\end{ruledtabular}
\end{table}

In this study, $e p \rightarrow e u_4 (d_4) X \rightarrow e q \gamma X$, $e p \rightarrow e u_4 (d_4) X \rightarrow q Z X \rightarrow e q \ell^+ \ell^- X$, $\gamma p \rightarrow u_4 (d_4) X \rightarrow q \gamma X$ and $\gamma p \rightarrow u_4 (d_4) X \rightarrow q Z X \rightarrow q \ell^+ \ell^- X$ processes (and their h.c.) are considered as  signatures of anomalous interactions of the fourth generation up and down type quarks ($q=u,c$ for $u_4$ and $q=d,s,b$ for $d_4$ and $\ell=e,\mu$). Since it is not possible to distinguish between $u_4$ and $d_4$, combined events are used in analysis.  The SM background for above processes is potentially much larger than the signal. However, after applying some kinematic cuts, it is posiible to decrease backgrounds to the reasonable levels. 

We choose the following two cut selection criteria for the first and third processes: the cut selection 1 criteria are $P_T\:>\:10$ GeV for the electron and photon, $P_T\:>\:20$ GeV for the jet, $\left|\eta_{j,\ell,\gamma}\right|\:<\:2.5$, $\Delta R_{j,\ell,\gamma}\:>\:0.4$ between the electron, photon and jet; the cut selection 2 criteria are $P_T\:>\:10$ GeV for the electron, $P_T\:>\:80$ GeV for photon, $P_T\:>\:20$ GeV for the jet, $\left|\eta_{j,e,\gamma}\right|\:<\:2.5$, $\Delta R_{j,\ell,\gamma}\:>\:0.4$ between the electron, photon and jet. The calculated signal and SM background cross sections for these processes with above selection cuts are given in Table I.

\begin{table}
\caption{Statistical significances for $q_4 \rightarrow \gamma q$ anomalous process at $ep$ and $\gamma p$ colliders with $\kappa_{g}=0$, $(\kappa_{\gamma}/\Lambda) = (\kappa_{Z}/\Lambda)=1$ TeV$^{-1}$ for cut Selection 2.}
\begin{ruledtabular}
\begin{tabular}{cccc}
 &\multicolumn{3}{c} {Statistical significances (SS)} \\
 & \multicolumn{2}{c}{$e p \rightarrow q_4 X\rightarrow  e \gamma j X+h.c.$}&$\gamma p \rightarrow q_4 X\rightarrow \gamma j X+h.c.$   \\
 \cline{2-3}  \cline{4-4}
 $m_{q_{4}}$ (GeV)&${L^{ep}}_{int}=10$ fb$^{-1}$&${L^{ep}}_{int}=4.1$ fb$^{-1}$&${L^{ep}}_{int}=4.1$ fb $^{-1}$\\
\hline
200 & 34.60 & 22.16 & 229.33  \\
300 & 22.65 & 14.54 & 173.97  \\
400 & 12.42 & 7.95 & 110.25  \\
500 & 6.38 & 4.08 & 66.27  \\
600 & 3.09 & 1.98 & 35.62 \\
700 & 1.35 & 0.86 & 16.68  \\
800 & 0.50 & 0.32 & 5.68  \\
\end{tabular}
\end{ruledtabular}
\end{table}

\begin{table}
\caption{Statistical significances for $q_4 \rightarrow Z q$ anomalous process at $ep$ and $\gamma p$ colliders with $\kappa_{g}=0$, $(\kappa_{\gamma}/\Lambda) = (\kappa_{Z}/\Lambda)=1$ TeV$^{-1}$ for cut Selection 2.}
\begin{ruledtabular}
\begin{tabular}{cccc}
 &\multicolumn{3}{c} {Statistical significances (SS)} \\
 & \multicolumn{2}{c}{$e p \rightarrow q_4 X\rightarrow e \ell^+ \ell^- j X+h.c.$}&$\gamma p \rightarrow q_4 X\rightarrow  \ell^+ \ell^- j X+h.c.$   \\
  \cline{2-3} \cline{4-4} 
 $m_{q_{4}}$ (GeV)&${L^{ep}}_{int}=10$ fb$^{-1}$&${L^{ep}}_{int}=4.1$ fb$^{-1}$&${L^{ep}}_{int}=4.1$ fb$^{-1}$\\
\hline
300 & 11.33 & 7.25 & 18.11 \\
400 & 12.26 & 7.85 & 42.60 \\
500 & 10.10 & 6.46 & 40.51  \\
550 & 6.87 & 4.40 & 29.82  \\
600 & 4.09 & 2.62 & 18.66  \\
700 & 2.07 & 1.33 & 9.33  \\
800 &  &  & 3.37 \\
\end{tabular}
\end{ruledtabular}
\end{table}

\begin{figure}[b]
\vspace{-0.2cm}
\subfigure[]{\includegraphics[width=7cm]{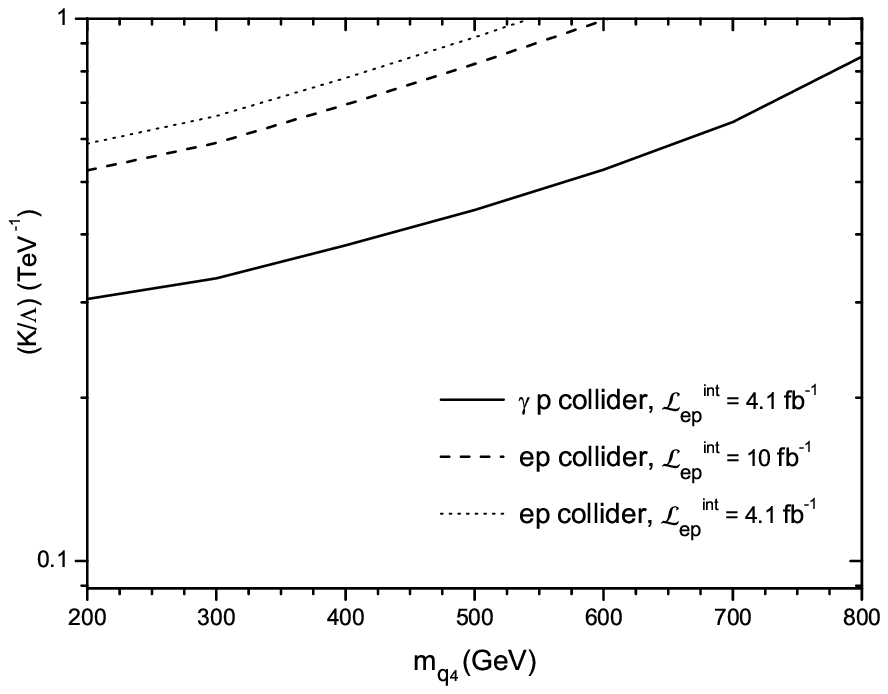}} \ \ \ \ \ \ \ \ \ \
\subfigure[]{\includegraphics[width=7cm]{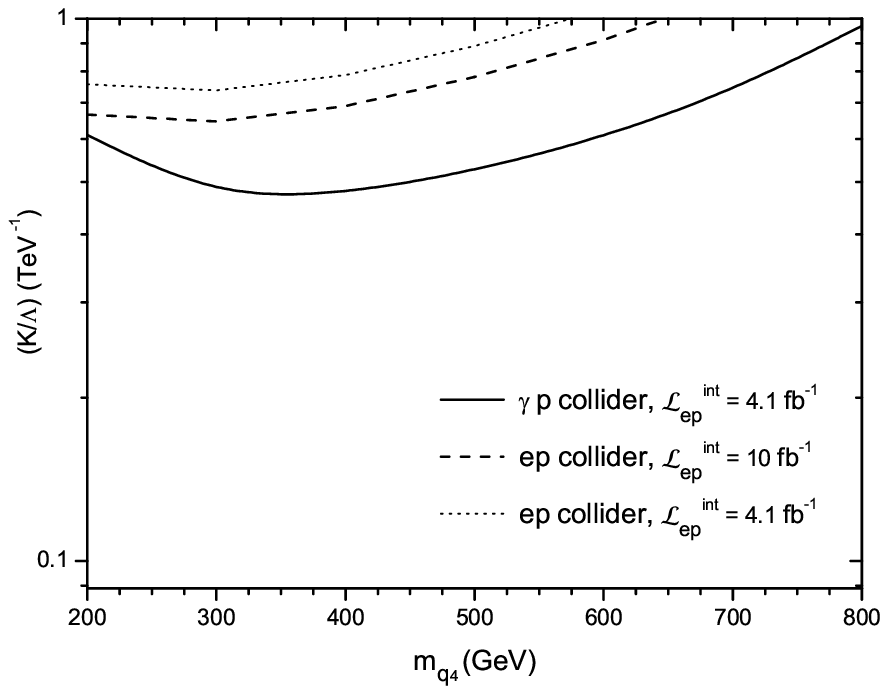}} \ \ \ \ \ \ \ \ \ \  
\subfigure[]{\includegraphics[width=7cm]{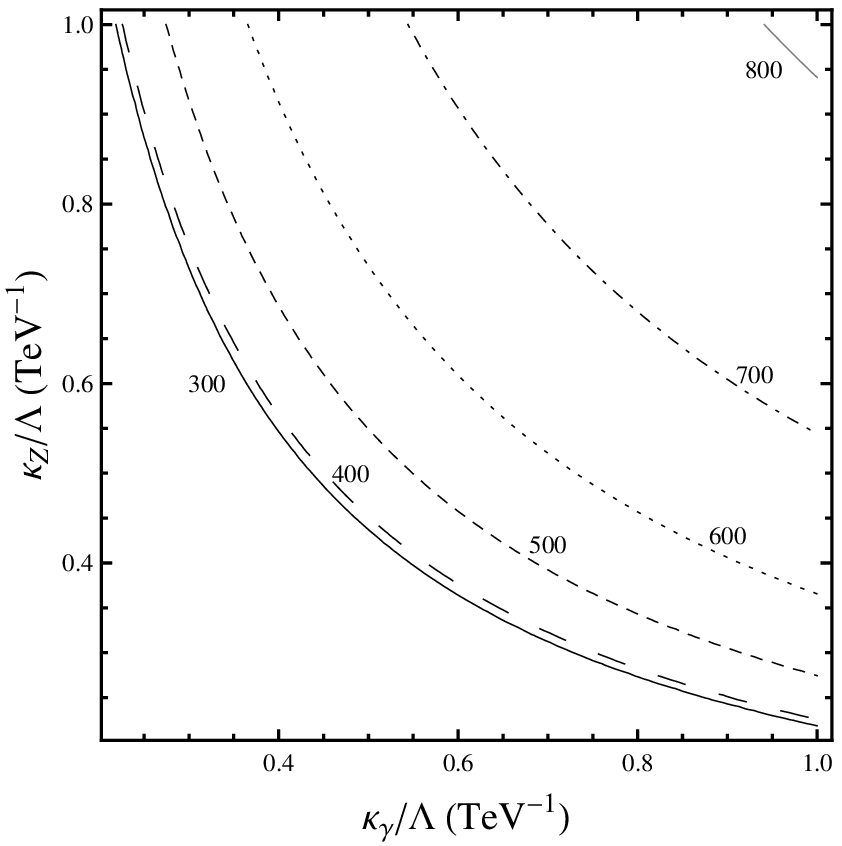}} 
\vspace{-0.2cm}
\caption{The achievable values of the anomalous coupling strength at $ep$ and $\gamma p$ colliders for a) $q_4 \rightarrow \gamma q$ anomalous process and (b) $q_4 \rightarrow Z q$ anomalous process as a function of the $q_{4}$ mass; (c) the reachable values of anomalous photon and Z couplings with $L_{int}=4.1$ fb$^{-1}$. \protect\label{fig4}}
\end{figure}

Similar but sligthly different cuts are applied for the second and fourth processes: the cut selection 1 criteria are $P_T\:>\:10$ GeV for the leptons, $P_T\:>\:20$ GeV for the jet, $\left|\eta_{j,\ell}\right|\:<\:2.5$, $\Delta R_{j,\ell}\:>\:0.4$ between leptons and the jet; the cut selection 2 criteria are $P_T\:>\:50$ GeV for leptons, $P_T\:>\:20$ GeV for the jet, $\left|\eta_{j,\ell,\gamma}\right|\:<\:2.5$, $\Delta R_{j,\ell}\:>\:0.4$ between the lepton and jet.  Again, the calculated signal and SM background cross sections for these processes with above selection cuts are given in Table II. 

The statistical significance (SS) values, evaluated from \cite{CMS}
\begin{equation}
SS=\sqrt{2 L_{int}\left[\left(\sigma_{S}+\sigma_B\right)ln\left(1+\frac{\sigma_{S}}{\sigma_B}\right)-\sigma_S \right]},
\end{equation} 
where $L_{int}$ is the integrated luminosity of the collider ($L_{int}^{max}=0.65 L_{int}^{ep}$ for the $\gamma p$ collider), are presented in Tables III  and IV  for the cut Selection 2.

One can compute achieveable values of the coupling strengths by setting both minimum of $SS$ to three and minimum of signal events to five. The results are presented at figure 4 for various fourth generation quark masses. The solid line at figure 4a corresponds to $\kappa_\gamma/\Lambda$ values independent of $\kappa_Z/\Lambda$. The rest of lines are for $\kappa/\Lambda=\kappa_\gamma/\Lambda=\kappa_Z/\Lambda$. When anomalous coupling strengts for electromagnetic and weak interactions are not equal, one can obtain them by using  $\gamma p \rightarrow u_4 (d_4) X \rightarrow q Z X \rightarrow q \ell^+ \ell^- X$ process (fig. 4c). Also, by using the relation for the statistical significance, it is possible to calculate the minimum luminosity to observe the considered processes depending on the fourth generation quark mass (figs. 5a and 5b) and the coupling strength for two different values of the mass (fig. 5c).   

\begin{figure}[b]
\vspace{-0.2cm}
\subfigure[]{\includegraphics[width=7cm]{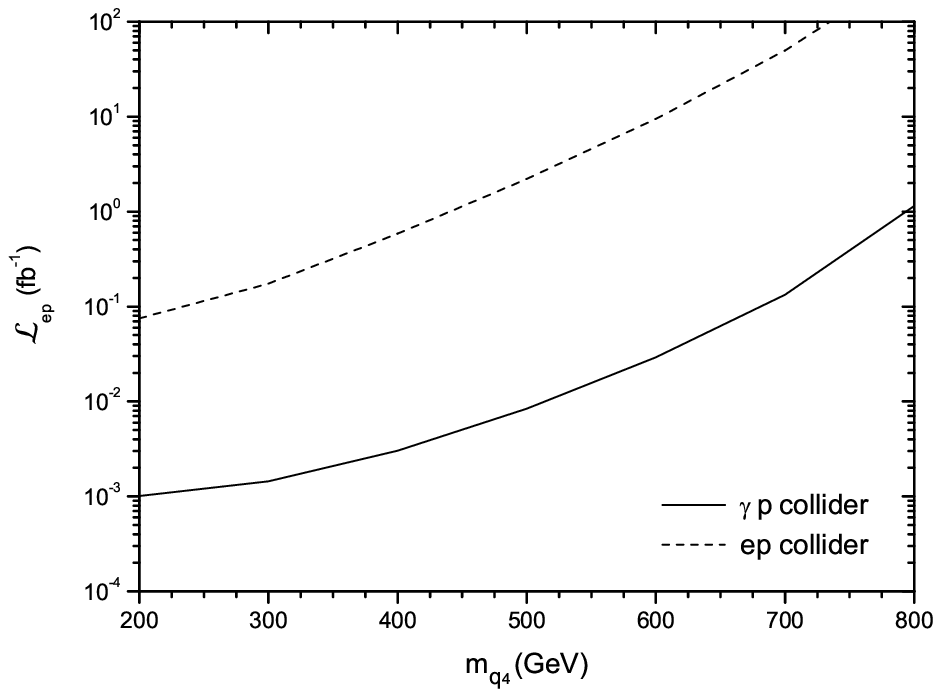}} \ \ \ \ \ \ \ \ \ \
\subfigure[]{\includegraphics[width=7cm]{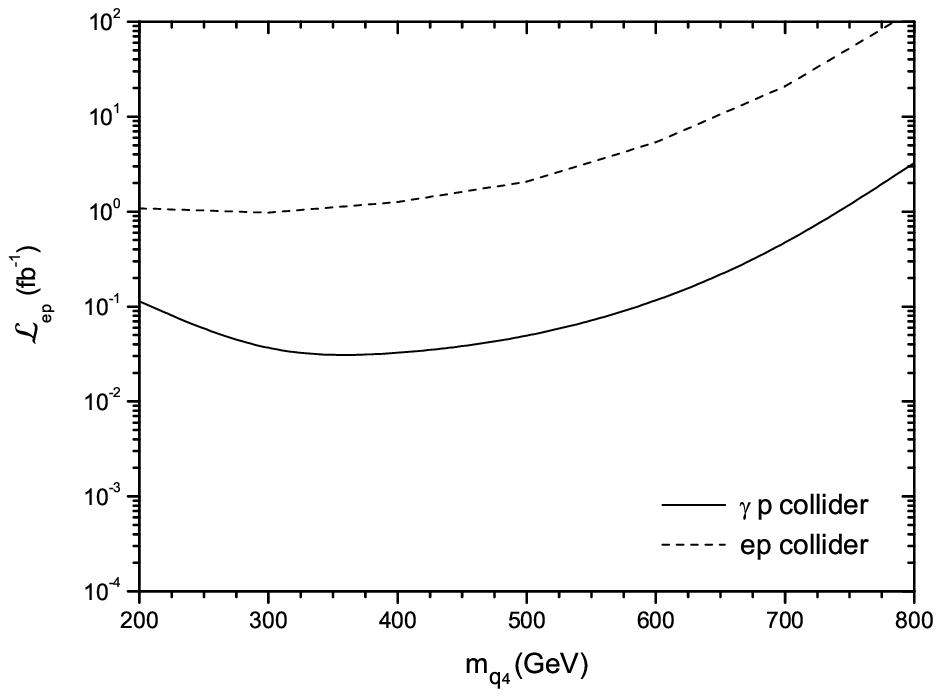}} \ \ \ \ \ \ \ \ \ \ 
\subfigure[]{\includegraphics[width=8cm]{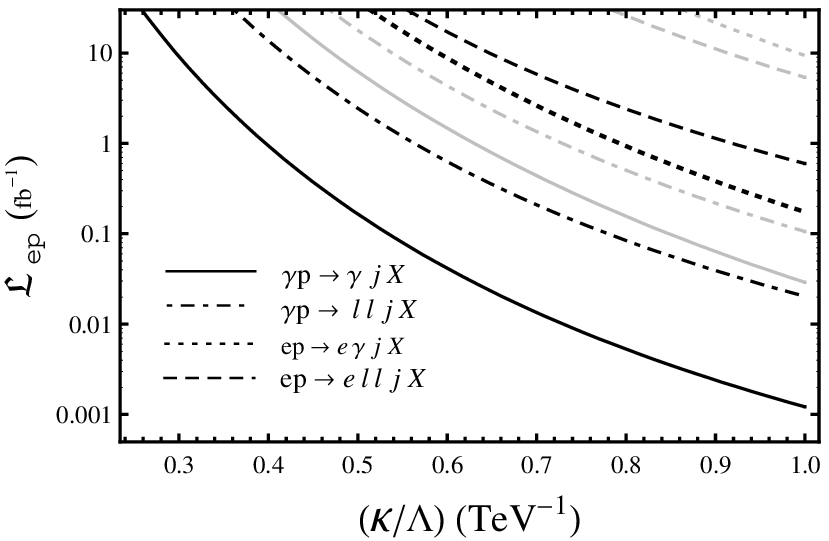}}
\vspace{-0.2cm}
\caption{The lowest necessary luminosity values of $ep$ colliders to observe (a) $q_4 \rightarrow \gamma q$ anomalous process and (b) $q_4 \rightarrow Z q$ anomalous process as a function of the $q_{4}$ mass; (c) as a function of anomalous coupling strength for $m_4 = 300$ GeV (black lines) and $m_4 = 600$ GeV (gray lines). \protect\label{fig5}}
\end{figure}

\section{Conclusion}

As a result of this study it is shown that when the anomalous coupling  for strong interactions is close to one, $ep$ and $\gamma p$ colliders are almost blind to anomalous interactions of the fourth generation quarks. In this case, hadron colliders like LHC will be discovery machines. However, $ep$ and $\gamma p$ colliders enable us to investigate effects of both anomalous couplings of electromagnetic and weak interactions for $\kappa^{q_{i}}_{g}=0$. This type anomalous interactions takes place through Weizsacker-Williams photons or Z boson at the LHC. Therefore, LHC based $ep$ and $\gamma p$ colliders have better observation reaches of anomalous couplings compare to the LHC.

Observation limits (at $3\sigma$) as low as $0.33$ ($0.52$) TeV$^{-1}$ are reachable for the $\kappa_\gamma/\Lambda$ at $\gamma p \rightarrow q_4 X\rightarrow \gamma j X+h.c.$ channel for fourth generation quarks with $m_4=300\: (600)$ GeV, regardless of the value of $\kappa_Z/\Lambda$. Even lower values can be reached at $\gamma p \rightarrow q_4 X\rightarrow  \ell^+ \ell^- j X+h.c.$ channel in case of either $\kappa_Z/\Lambda$ or $\kappa_\gamma/\Lambda$ equals to 1 TeV$^{-1}$. In this situation, the low values for the other coupling strength become 0.22 TeV$^{-1}$ and 0.36 TeV$^{-1}$ for $q_4$ with $300$ and $600$ GeV masses, respectively. Meanwhile, $0.59$ ($0.91$) TeV$^{-1}$ for the $\kappa/\Lambda$ is reachable at ep collider with $L_{int}=10\:fb^{-1}$ for $m_4=300\: (600)$ GeV. The advantage of the $\gamma p$ collider with respect to $ep$ collider is obvious even with a quarter of luminosity of ep collider. 

\begin{acknowledgments}

We would like to thank S. Sultansoy and F. Zimmermann for many helpful conversations and discussions. This work is supported by Turkish Atomic Energy Authority (TAEA) via "$ep$, $\gamma p$, $eA$ and $\gamma A$ colliders on the bases of the Linac-LHC " project and Turkish State Planning Organization (DPT) with grant number DPT2006K-120470. 
\end{acknowledgments}

\end{document}